\documentclass[prd,a4paper,eqsecnum,nofootinbib,preprint,tightenlines]{revtex4}
\usepackage{epsfig}
\usepackage{bm}

\def\ba{{\bm{a}}}
\def\bk{{\bm{k}}}
\def\bn{{\bm{n}}}
\def\bmm{{\bm{m}}}

\def\muhat{\hat{\bm{\mu}}}
\def\Tr{\text{Tr}\,}

\def\Det{\text{Det}\,}
\def\Eq#1{Eq.~(\ref{#1})}
\def\nnn{{\em nnn\/}}
\renewcommand\Im{\text{Im}\,}

\begin{document}
\title{Lattice gauge theory with baryons at strong coupling}
\author{Barak Bringoltz}
\author{Benjamin Svetitsky}
\affiliation{School of Physics and Astronomy, Raymond and Beverly
Sackler Faculty of Exact Sciences, Tel Aviv University, 69978 Tel
Aviv, Israel}
\begin{abstract}
We study the effective Hamiltonian for strong-coupling lattice QCD
in the case of non-zero baryon density.
In leading order the effective Hamiltonian is a generalized antiferromagnet.
For naive fermions, the symmetry is $U(4N_f)$ and the spins belong
to a representation that depends on the local baryon number.
Next-nearest neighbor (\nnn) terms in the Hamiltonian break the symmetry to
$U(N_f)\times U(N_f)$.
We transform the quantum problem to a Euclidean sigma model which
we analyze in a $1/N_c$ expansion.
In the vacuum sector we recover spontaneous breaking
of chiral symmetry for the nearest-neighbor and \nnn\ theories.
For non-zero baryon density we study the nearest-neighbor theory only,
and show that the pattern of spontaneous symmetry breaking depends on
the baryon density.
\end{abstract}
 \maketitle

\section{Introduction}

The study of quantum chromodynamics at high density is almost as
old as the theory itself \cite{Collins}. In recent years the field
has attracted wide interest in the wake of a revival of the idea
of color superconductivity (CSC) \cite{Barrois,Bailin}. The
stimulus for this revival was the observation \cite{Alford,Rapp}
that the instanton-induced quark--quark interaction can be much
stronger than that induced by simple one-gluon exchange, and can
thus give a transition temperature on the order of 100~MeV.
Subsequent work \cite{Son} showed that the perturbative
color-magnetic interaction also gives rise to a strong pairing
interaction.

These and other dynamical considerations \cite{SchaferWilczek}
underlie a picture of the ground state of high-density QCD in
which the $SU(3)$ gauge symmetry is spontaneously broken by a
BCS-like condensate. The details of the breaking, which include
both the Higgs (or Meissner) effect and the rearrangement of
global symmetries and Goldstone bosons, depend on quark masses,
chemical potentials, and temperature. Prominent in the list of
possibilities are those of color-flavor locking in three-flavor
QCD \cite{Alford1} and crystalline superconductivity---with broken
translation invariance---when there are two flavors with different
densities \cite{Alford2}. For a review see \cite{Krishna}.

As noted, CSC at high density is so far a prediction of
weak-coupling analysis. One expects the coupling to become weak
only at high densities, and in fact it turns out that reliable
calculations demand extremely high densities \cite{Shuster}. The
use of weak-coupling methods to make predictions for moderate
densities is thus not an application of QCD, but of a model based
on it.  It is imperative to confirm these predictions by
non-perturbative methods.  Standard lattice Monte Carlo methods,
unfortunately, fall afoul of well-known technical problems when
the chemical potential is made nonzero, although we do note
remarkable progress made recently in the small-$\mu$ regime
\cite{Fodor,Owe}.

In this paper we initiate a study of high-density quark matter
based on lattice QCD in the strong-coupling limit.\footnote{ An
early discussion of our program, with early results, was given in
\cite{Boston}.} We work in the Hamiltonian formalism, which is
more amenable than the Euclidean formalism to strong-coupling
perturbation theory and to qualitative study of the ensuing
effective theory \cite{BSK,SDQW,Smit,BS}.  The fermion kinetic
Hamiltonian is a perturbation that mixes the zero-flux states that
are the ground-state sector of the electric term in the gauge
Hamiltonian. In second order, it moves color-singlet fermion pairs
around the lattice; the effective Hamiltonian for these pairs is a
generalized antiferromagnet, with spin operators constructed of
fermion bilinears.

We depart from studies of the vacuum by allowing a background baryon
density, which is perforce static in second order in perturbation
theory.  Our aim at this stage is to discover the ground state of
the theory with this background.  In third order (when $N_c=3$)
the baryons become dynamical; we display the effective Hamiltonian but make
no attempt to treat it.

The symmetry group of the effective antiferromagnet is the same as
the global symmetry group of the original gauge theory.  This
depends on the formulation chosen for the lattice fermions.
Following \cite{SDQW}, we begin with naive, nearest-neighbor
fermions, which suffer from species doubling \cite{NN} and possess
a global $U(4N_f)$ symmetry group that contains the ordinary
chiral symmetries [as well as the axial $U(1)$]  as subgroups.  We
subsequently break the too-large symmetry group with
next-nearest-neighbor (\nnn) couplings along the axes in the
fermion hopping Hamiltonian.  A glance at the menu of fermion
formulations reveals the reasons for our choice.  Wilson fermions
\cite{Wilson} have no chiral symmetry and make comparison of
results to continuum CSC difficult if not impossible.  Staggered
fermions \cite{Susskind} likewise possess only a reduced axial
symmetry while suffering a reduced doubling problem.  The overlap
action \cite{overlap} is non-local in time and hence possesses no
Hamiltonian; attempts \cite{Creutz} to construct an overlap
Hamiltonian directly have not borne fruit.  Finally, domain-wall
fermions \cite{Kaplan,Shamir} have been shown \cite{BS} to lose
chiral symmetry and regain doubling when the coupling is strong.

As we discuss below, while the \nnn\ theory still exhibits
doubling in the free fermion spectrum, we are not interested in
the perturbative fermion propagator but in the spectrum of the
confining theory. We take it as a positive sign that the unbroken
symmetry is now $U(N_f) \times U(N_f)$. This symmetry is what we
want for the continuum theory, except for the axial $U(1)$.  The
latter can still be broken by hand.\footnote{The breaking of the
naive fermions' symmetry by longer-range terms is a feature
\cite{SDQW} of SLAC fermions \cite{DWY} and also occurs if naive
fermions are placed on a {\em bcc\/} lattice \cite{Myint}.}

Our emphasis on the global symmetries is a consequence of the fact
that the gauge field is not present in the ground-state sector and
does not reappear in strong-coupling perturbation theory.  In
other words, confinement is a {\em kinematic\/} feature of the
theory, leaving no possibility of seeing the Higgs-Meissner effect
directly.  This is but an instance of confinement-Higgs duality,
typical of gauge theories with matter fields in the fundamental
representation \cite{duality}.  Our aim is thus to identify the
pattern of spontaneous breaking of global symmetries.  For various
values of $N_c$ and $N_f$, this can be compared to weak-coupling
results \cite{Schafer}.

This paper is largely an exposition of formalism, along with
partial results. We study the nearest-neighbor antiferromagnetic
Hamiltonian, both with and without a uniform baryonic background
density. We transform the quantum Hamiltonian into a path integral
for a non-linear $\sigma$ model, where the manifold of the
$\sigma$ field depends on the baryon background. We then
investigate the limit of large $N_c$ and show that the global
$U(4N_f)$ symmetry is indeed spontaneously broken.

Adding in the \nnn\ couplings is a problem of vacuum alignment
\cite{Peskin}. We do this in the vacuum sector and recover the
result \cite{SDQW} that the $U(N_f)\times U(N_f)$ chiral symmetry
is broken to the vector $U(N_f)$.  The analysis for the
finite-density theory is more involved and we defer it to a future
publication.

Other groups have recently studied the strong-coupling effective
Hamiltonian for naive and Wilson fermions
at non-zero chemical potential \cite{Yaouanc,Gregory,Umino}.
We differ from their approaches in eschewing mean field theory in favor
of the exact transformation to the $\sigma$ model, which is amenable
to semiclassical treatment. As noted above, we base our program on
\nnn\ fermions; we also work at fixed baryon density.

Let us walk through the paper. We review in Sec.~\ref{sec:Heff}
the derivation of the effective Hamiltonian of lattice gauge
theory in strong-coupling perturbation theory \cite{SDQW,Smit}.
The second-order Hamiltonian [$O(1/g^2)$] is an antiferromagnet
with $U(4N_f)$ spins; the global symmetry group is $U(4N_f)$ for
the nearest-neighbor theory, broken to $U(N_f)\times U(N_f)$ by
\nnn\ terms. The baryon number at each site determines the
representation of $U(4N_f)$ carried by the spin at that site. In
second order, baryon number is static; it becomes mobile in the
next order, where (for $N_c=3$) the new term in the effective
Hamiltonian is a baryon hopping term.

The baryon operators responsible for the hopping are composite
operators that do not obey canonical anti-commutation relations.
If this were not the case, then the effective Hamiltonian in third
order would strongly resemble that of the $t$--$J$ model
\cite{Auerbach},
\begin{equation}
H_{\text{$t$--$J$}}=-t\sum_{\langle ij\rangle\atop
s}c^{\dag}_{is}c_{js} +J\sum_{\langle ij\rangle}\left(\bm
S_i\cdot\bm S_j-\frac{n_in_j}4\right)+\mathcal J'.
\end{equation}
Here $c_{js}$ is an annihilation operator for an electron at site
$j$ with spin $s$, and the number operators $n_i=c^\dag_ic_i$ and
spin operators $\bm S_i=\frac12c^\dag_i\bm\sigma c_i$ are
constructed from it. The added term $\mathcal J'$ is a more
complicated hopping and interaction term.  The $t$--$J$ model
describes a doped antiferromagnet; it arises as the strong-binding
limit of Hubbard model, a popular model for itinerant magnetism
and possibly for high-$T_c$ superconductivity. The model is not
particularly tractable and, absent new theoretical developments,
does not offer much hope for progress in our finite-density
problem. It is nonetheless worth pondering the fact that a model
connected, however tentatively, with superconductivity appears in
a study of high-density nuclear matter.

\begin{figure}[htb]
        \epsfig{width=6cm,file=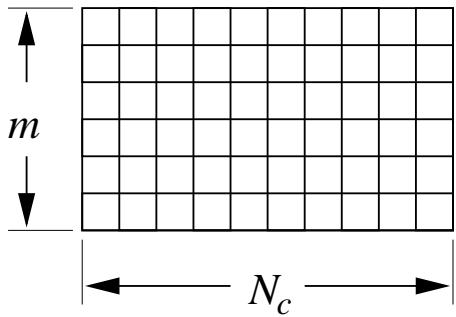} \caption{The
        representation of $U(4N_f)$ carried by the spin in the
        effective antiferromagnet.  $m$ is related to the baryon
        number at the site according to $m=B+2N_f$, with
        $|B|\leq 2N_f$.}\label{fig:young1}
\end{figure}

In the remainder of this paper, we work only to $O(1/g^2)$, where
the baryons are fixed in position. Motivated by the similarity of
our Hamiltonian to the Heisenberg antiferromagnet, we apply
condensed matter methods developed for that problem. Indeed,
condensed matter physicists have generalized the $SU(2)$,
spin-$1/2$ Heisenberg model to $SU(N)$ in many representations
\cite{Haldane,Affleck,MA,AA,RS1,RS2,Salam,Auerbach}, which
corresponds to adding flavor and color degrees of freedom to the
electrons.\footnote{We refer the reader to the paper by Read and
Sachdev \cite{RS1} for a survey, including a phase diagram in the
$(N,N_c)$ plane.} These are exactly the generalizations needed for
our effective Hamiltonian. With $N_c$ colors and $N$
(single-component) flavors, a site of the lattice can be
constrained to contain a color-singlet combination of $mN_c$
particles. The flavor indices of the spin then make up a
representation of $SU(N)$ whose Young diagram has $N_c$ columns
and $m$ rows (see Fig.~\ref{fig:young1}). We set
\begin{equation}
N=4N_f
\end{equation}
and the correspondence is complete (until we include \nnn\ terms
in the Hamiltonian).

In Sec.~\ref{sec:NLS} we derive a $\sigma$ model representation
for the partition function of the antiferromagnet.  Following Read
and Sachdev \cite{RS1}, we employ spin coherent states
\cite{Klauder} to define the $\sigma$ field.  $N$ and $m$
determine the target space of the $\sigma$ model to be the
symmetric space $U(N)/[U(m)\times U(N-m)]$; the number of colors
$N_c$ becomes an overall coefficient of the action.\footnote{The
inverse gauge coupling $1/g^2$ multiplies the quantum Hamiltonian,
and hence serves only to set the energy scale.} As for the quantum
Hamiltonian, the nearest-neighbor theory is symmetric under $U(N)$
while the \nnn\ terms break the symmetry to $U(N_f)\times U(N_f)$
(while leaving the manifold unchanged).

The $N_c$ multiplying the action invites a large-$N_c$ analysis,
and in Sec.~\ref{sec:zeroB} we study the vacuum sector, meaning
zero baryon number, thereby.  We return to an exercise proposed
and solved by Smit \cite{Smit}, in generalizing the vacuum sector
to allow baryon number $\pm B$ on alternating sites; this means
specifying conjugate representations of $U(N)$ on alternating
sites, with respectively $m$ and $N-m$ rows.  As shown by Read and
Sachdev \cite{RS1}, in this situation one can carry out an
alternating $U(N)$ rotation to convert the antiferromagnet into a
ferromagnet with {\em identical\/} spins on alternating sites, and
the classical ($N_c=\infty$) analysis gives a homogeneous ground
state. The result is, as one might expect, that $U(N)$ is broken
to $U(m)\times U(N-m)$ in the classical vacuum; the ground state
energy is independent of $m$. The $1/N_c$ corrections to the
energy do depend on $m$, however, and they select the
self-conjugate $m=N/2$ configuration ({\em i.e.}, $B=0$
everywhere) as the lowest-energy vacuum.  Thus the true ground
state breaks $U(N)\to U(N/2)\times U(N/2)$.\footnote{This result
was obtained by Smit using a Holstein-Primakoff transformation on
the quantum Hamiltonian.  We note in passing that the $1/N_c$
calculation includes the effect of the time-derivative terms in
the action that were dropped in the leading order, and these terms
{\em do\/} remember the difference between the ferromagnet and the
antiferromagnet.}  When we add \nnn\ terms to the action as a
perturbation, we find that the ground state breaks $U(N_f)\times
U(N_f)$ to the vector $U(N_f)$, as expected.

We turn to non-zero baryon density in Sec.~\ref{sec:nonzeroB}. We
study homogeneous states, in which all sites carry the same
representation of $U(N)$, with $m>N/2$. The classical vacuum of
the $\sigma$ model is more elusive than for the vacuum sector,
since now there are identical manifolds on adjacent sites but the
coupling is antiferromagnetic.  We begin by studying the two-site
problem, and we learn that when one of the classical spins is
fixed then when the energy is minimized the other spin is still
free to wander a submanifold of the original symmetric space.  If
we replicate this to the infinite lattice then we have a situation
where the even spins, say, are fixed in direction while each odd
spin wanders the submanifold, independent of the other odd spins.
This means a ground state whose degeneracy is exponential in the
volume, similar to some frustrated models or the antiferromagnetic
Potts model \cite{AFPotts}.  The cure to this disease comes from
the $O(1/N_c)$ fluctuations, which couple the odd spins to each
other and make them align.  In the end we find that the $U(N)$
symmetry is broken by the vacuum to $U(2m-N)\times U(N-m)\times
U(N-m)$. Perturbing this ground state with the \nnn\ terms is
technically difficult, and we do not attempt it here despite its
obvious physical interest.

We close with a brief summary and discussion.
The $O(1/N_c)$ calculation in the $B\not=0$ case is
relegated to an appendix, as are other (but not all) technical
details.

\section{The effective Hamiltonian \label{sec:Heff}}

For an $SU(N_c)$ gauge theory with $N_f$ flavors of fermions, we
write the lattice Hamiltonian
\begin{equation}
H=H_E+H_U+H_F.
\end{equation}
Here $H_E$ is the electric term, a sum over links $(\bn\mu)$ of
the $SU(N_c)$ Casimir operator on each link,
\begin{equation}
H_E=\frac12g^2\sum_{\bn\mu}E_{\bn\mu}^2.
\end{equation}
Next is the magnetic term, a sum over plaquettes,
\begin{equation}
H_U=\frac1{2g^2}\sum_p\left(N_c-\Tr U_p\right).
\end{equation}
Finally we have the fermion Hamiltonian,
\begin{equation}
H_F=-i\sum_{\bn\mu}\psi^{\dag\alpha f}_\bn\alpha_\mu
\sum_{j>0}D(j)\left(\prod_{k=0}^{j-1}U_{\bn+k\muhat,\mu}\right)
_{\alpha\beta} \psi^{\beta f}_{\bn+j\muhat}+\text{\em h.c.}
\end{equation}
The fermion field $\psi^{\alpha f}_\bn$ carries color $\alpha$ and
flavor $f$ at site $\bn$. The function $D(j)$ is a kernel that
defines the lattice fermion derivative. It can yield a naive,
nearest-neighbor action if $D(j)=\frac12\delta_{j,1}$; a
long-range SLAC derivative \cite{DWY} if $D(j)=-(-1)^j/j$; or
anything in between, such as a \nnn\ action obtained by truncating
the SLAC kernel to its $j=1,2$ terms.

For $g\gg1$ the ground state of $H$ is determined by $H_E$ alone
to be any state with zero electric field, whatever its fermion
content,
\begin{equation}
|0\rangle|\chi\rangle_F=\left[\prod_{\bn\mu}|E_{\bn\mu}^2=0\rangle\right]
|\chi\rangle_F.
\end{equation}
These states have energy $\epsilon_0=0$ and are degenerate with
respect to all the fermionic degrees of freedom. We consider
perturbation theory in $V=H_U+H_F$. Both $H_U$ and $H_F$ are sums
of operators that are strictly bounded, independent of $g$ except
for the explicit coefficient in $H_U$. We can dismiss first-order
perturbations by noting that $H_U$ and $H_F$ are multilinear in
link operators $U$ and $U^\dag$, which are raising/lowering
operators for the electric field; thus there are no non-zero
matrix elements within the zero-field sector.

We proceed to higher orders, and seek an effective Hamiltonian
that acts in the zero-field sector \cite{Kato}. Define $P_0$ to be
the projector onto the subspace of all the $E=0$ states. Then
perturbation theory in $V$ gives an effective Hamiltonian,
\begin{equation}
H_{\text{eff}}=P_0VQDVP_0+P_0VQDVQDVP_0+\cdots.
\end{equation}
Here $Q=1-P_0$ projects onto the subspace orthogonal to the $E=0$
states; the operator $D\equiv (\epsilon_0-H_E)^{-1}$ supplies
energy denominators, so that
\begin{equation}
QD=\sum_{\text{$E\not=0$
states}}|\lambda\rangle\frac1{\epsilon_0-\epsilon_\lambda}\langle
\lambda|.
\end{equation}

The intermediate states $|\lambda\rangle$ contain flux
excitations. In second and third order the patterns of flux can
only be strings of length $j$ in the fundamental representation of
the color group. Thus the energy denominators are
\begin{equation}
\epsilon_0-\epsilon_\lambda=-\frac12g^2C_F|j|,
\end{equation}
where $C_F=(N_c^2-1)/(2N_c)$ is the quadratic Casimir of the
fundamental representation of $SU(N_c)$.

The perturbations $H_U$ and $H_F$ are explicitly of $O(1/g^2)$ and
$O(1)$, respectively; each energy denominator gives a factor of
$1/g^2$. Thus to $O(1/g^4)$ we can forget about $H_U$. Our result
to this order is
\begin{equation}
H_{\text{eff}}=P_0H_FDH_FP_0+P_0H_FDH_FDH_FP_0. \label{Heff}
\end{equation}
Since $H_F$ has no non-zero matrix elements within the $E=0$
sector, we have dispensed with $Q$ in \Eq{Heff}. The first term in
\Eq{Heff} arises for any value of $N_c$ and is $O(1/g^2)$; the
case $N_c=2$ must be treated carefully, but all cases $N_c>2$ are
generic. The second term is special to $N_c=3$ and is $O(1/g^4)$.

\subsection{Second order: the antiferromagnet}

We calculate explicitly the first term in $H_{\text{eff}}$. Each
term in $H_F$ creates a string of flux of length $j$, which must
be destroyed by the conjugate term. Thus
\begin{equation}
H_{\text{eff}}^{(2)}=2\sum_{j>0} [-K(j)]\sum_{\bn\mu}
\left(\psi_\bn^{\dag\alpha f}\alpha_\mu\psi_{\bn+j\muhat}^{\beta
f}\right) \langle0|\left(\prod U\right)_{\alpha\beta} \left(\prod
U^\dag\right)_{\gamma\delta}|0\rangle
\left(\psi_{\bn+j\muhat}^{\dag\gamma g}\alpha_\mu\psi_\bn^{\delta
g}\right), \label{H2eff1}
\end{equation}
where we define
\begin{equation}
K(j)= \frac{\left[D(j)\right]^2} {\frac12g^2C_F|j|}>0.
\end{equation}
The matrix element of the gauge fields yields
$\frac1{N_c}\delta_{\alpha\delta}\delta_{\beta\gamma}$,
independent of $j$.

As they appear in \Eq{H2eff1}, each $\psi^\dag$ is next to a
$\psi$ on a different site. This invites a Fierz transformation on
the product of fermion fields, which we write generally as
\begin{equation}
\left(\psi^{\dag}_i\alpha_\mu\psi_j\right)\left(\psi^{\dag}_k\alpha_\mu\psi_l\right)
= \delta_{jk}\psi^{\dag}_i\psi_l -\frac14\sum_{A}s_{A}^{\mu}
\left(\psi^{\dag}_i\Gamma^A\psi_l\right)
\left(\psi^{\dag}_k\Gamma^A\psi_j\right). \label{befFierz}
\end{equation}
Here $i,j,k,l$ are combined site, flavor, and color indices, and
we have assumed that $k$ and $l$ are always different while $j$
and $k$ might be equal [as in \Eq{H2eff1}]. The matrices
$\Gamma^A$ are the 16 Dirac matrices, normalized to
$(\Gamma^A)^2=\bm1$, and we have defined
\begin{equation}
s_{A}^{\mu}=\frac14\Tr\Gamma^A\alpha_\mu\Gamma^A\alpha_\mu=\pm1.
\end{equation}
This sign factor is $\pm1$ according to whether $\Gamma^A$
commutes or anticommutes with $\alpha_\mu$; it will be a constant
companion in our calculations. As they appear in
$H_{\text{eff}}^{(2)}$, the indices $i,l$ are the same site and
color but different flavors, and likewise $j,k$. Leaving the
flavor indices explicit, we obtain
\begin{equation}
H_{\text{eff}}^{(2)}= \frac1{4N_c}\sum_{\bn\mu\atop j\not=0} K(j)
s_A^\mu \left(\psi^{\dag f}\Gamma^A\psi^g\right)_\bn
\left(\psi^{\dag g}\Gamma^A\psi^f\right)_{\bn+j\muhat}
-d\sum_{j\not=0}K(j) \sum_\bn\left(\psi^{\dag f}\psi^f\right)_\bn.
\end{equation}
Each fermion bilinear in parentheses is a color singlet located at
a given site. The second term contains the baryon
density\footnote{ This baryon number is positive semidefinite, and
is zero for the drained state (see below). The conventional baryon
number $B_\bn$ is zero in the half-filled state, and thus
$B_\bn'=B_\bn+2N_f$.} $B'_\bn=N_c^{-1}\psi^{\dag}_\bn\psi_\bn$,
and the sum $\sum_\bn B'_\bn$ is the total baryon number $B'$.

We now combine the Dirac indices with the flavor indices and write
\begin{equation}
\left(\psi^{\dag f}\Gamma^A\psi^g\right)_\bn \left(\psi^{\dag
g}\Gamma^A\psi^f\right)_{\bn'} =8\left(\psi^\dag
M^\eta\psi\right)_\bn\left(\psi^\dag M^\eta\psi\right)_{\bn'}.
\end{equation}
We have defined new matrices $M^\eta$ as direct products of the
$4\times4$ Dirac matrices and the $U(N_f)$ flavor generators,
\begin{equation}
M^\eta=\Gamma^A\otimes\lambda^a, \label{dirprod}
\end{equation}
and we have normalized them conventionally according to
\begin{equation}
\Tr M^\eta M^{\eta'}=\frac12\delta^{\eta\eta'}.
\end{equation}
The $M^\eta$ generate a $U(N)$ algebra, with $N\equiv4N_f$.

An alternating flip
\begin{equation}
\psi_\bn\to \left[\prod_\mu (\alpha_\mu)^{n_\mu}\right]\psi_\bn
\end{equation}
(spin diagonalization \cite{SDQW}) removes the $\alpha_\mu$
matrices from the odd-$j$ terms in $H_F$, and hence removes the
sign factors $s_A^\mu$ from the odd-$j$ terms in
$H_{\text{eff}}^{(2)}$. We have finally
\begin{equation}
H_{\text{eff}}^{(2)}= \frac2{N_c}\sum_{\bn\mu j} K(j)
\left(s_\eta^\mu\right)_{\text{even $j$}\atop\text{only}}
\left(\psi^{\dag}M^\eta\psi\right)_\bn
\left(\psi^{\dag}M^\eta\psi\right)_{\bn+j\muhat}
-\left(dN_c\sum_jK(j)\right)B'. \label{H2eff2}
\end{equation}
The odd-$j$ terms are of the form $\bm{M}\cdot\bm{M}$ which can be
written in any basis for the $U(4N_f)$ algebra. The even-$j$
terms, however, contain $s_\eta^\mu$ which is defined only in the
original basis (\ref{dirprod}).

\subsection{Single-site states}

In the zero-field sector in which we work, Gauss' Law constrains
the fermion state at each site to be a color singlet. The drained
state $|\text{dr}\rangle$, with $\psi^{\alpha
f}_\bn|\text{dr}\rangle=0$ for all $(\alpha,f)$, is the unique
state with $B'=0$. The other color singlet states may be generated
by repeated application of the baryon creation operator,
\begin{equation}
b^\dag_{fg\cdots}= \epsilon_{\alpha\beta\cdots}\psi^{\dag \alpha
f}_\bn \psi^{\dag \beta g}_\bn\cdots,
\end{equation}
with $N_c$ operators $\psi$. (Here and henceforth, the indices
$f,g,\ldots$ combine the flavor and Dirac indices.)

\begin{figure}[htb]
        \epsfig{width=2cm,file=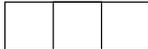}
        \caption{Young diagram of the representation of $U(4N_f)$
carried by the baryon operator}
        \label{fig:baryon}
\end{figure}

As noted above, at each site $\bn$ the operators
\begin{equation}
Q^\eta_\bn=\psi^\dag_\bn M^\eta \psi_\bn = \psi^{\dag\alpha f}_\bn
M^\eta_{fg}\psi_\bn^{\alpha g} \label{fermionrep}
\end{equation}
generate a $U(N)$ algebra, with $N=4N_f$. The drained state is
obviously a singlet under this algebra. The creation operator
$b^\dag_{fg\cdots}$ is in the symmetric representation of $U(N)$
with one row and $N_c$ columns (see Fig.~\ref{fig:baryon}).
Repeated application of $b^\dag_{fg\cdots}$ to the drained state
gives the state
\begin{equation}
|\chi\rangle=b^\dag b^\dag \cdots|\text{dr}\rangle.
\end{equation}
If there are $m$ operators $b^\dag$, then the state $|\chi\rangle$
lies in the representation with $N_c$ columns and $m$ rows (see
Fig.~\ref{fig:young1}). Its baryon number is $B'_\bn=m$.

The second-order effective Hamiltonian $H_{\text{eff}}^{(2)}$
preserves $B'_\bn$, the baryon number on each site. Thus any
distribution of $B'_\bn$ defines a sector within which
$H_{\text{eff}}^{(2)}$ is to be diagonalized. In other words,
baryons constitute a fixed background in which to study
``mesonic'' dynamics. The baryon number at each site fixes the
representation of $U(N)$ at that site, which is the space of
states in which the charges $Q^\eta_\bn$ act.

\subsection{Global symmetries and doubling}

The $j=1$ terms in \Eq{H2eff2} are of the form $Q^\eta_\bn
Q^\eta_{\bn+\muhat}$, and they commute with the generators
\begin{equation}
Q^\eta=\sum_\bn Q^\eta_\bn
\end{equation}
of a global $U(N)$ symmetry group. This symmetry is in fact
familiar from the lattice Hamiltonian of naive, nearest-neighbor
fermions: Spin diagonalization of $N_f$ naive Dirac fermions
transforms the Hamiltonian into that of $4N_f$ staggered fermions.
In the weak coupling limit, there are in fact $8N_f$ fermion
flavors---the doubling problem. This doubling is partially
reflected in the accidental $U(4N_F)$ symmetry, which is intact in
the $g\to\infty$ limit and is respected by the effective
Hamiltonian. Retaining terms in the fermion Hamiltonian (and thus
in $H^{(2)}_{\text{eff}}$) that involve odd separations $j$ does
not break this symmetry.

The Nielsen-Ninomiya theorem \cite{NN} guarantees that any fermion
Hamiltonian of finite range will possess the full doubling
problem. This is a statement, however, about weak coupling only,
since the dispersion relation of free fermions is irrelevant if
the coupling is strong and the fermions are confined. It is
interesting that the accidental $U(4N_f)$ symmetry nonetheless
survives into strong coupling as a vestige of doubling.

The terms in \Eq{H2eff2} with {\em even\/} $j$, on the other hand,
break the $U(N)$ symmetry, as do even-$j$ terms in the original
fermion Hamiltonian. It is easy to see via spin diagonalization,
which leaves the even-$j$ terms unchanged, that the only
generators left unbroken are the $Q^\eta$ corresponding to
\begin{equation}
M^\eta=\bm{1}\otimes\lambda^a\quad\text{and}\quad
\gamma_5\otimes\lambda^a,
\end{equation}
which form the $U(N_f)_L\times U(N_f)_R$ chiral algebra. This of
course makes no difference to the Nielsen-Ninomiya theorem, which
will enforce 8-fold doubling in the perturbative propagator even
without the $U(4N_f)$ symmetry. If we are interested in the
realization of the global symmetries of the continuum theory,
though, we can study this lattice theory which has the same
symmetry. The simplest theory one may study is thus one containing
nearest-neighbor and \nnn\ terms. We shall proceed to discard
terms with longer range; we shall begin with the nearest-neighbor
theory, with its accidental doubling symmetry, and later break
this symmetry to $U(N_f)_L\times U(N_f)_R$ with the \nnn\ terms.

Two essential differences will always remain between this lattice
theory and the continuum theory. One is the presence of the axial
$U(1)$ symmetry on the lattice. This symmetry is exact, broken by
no anomaly, and may make the drawing of conclusions for the
continuum theory less than straightforward unless it is broken by
hand. The other difference is the fact that the effective
Hamiltonian for baryons (see below) is also a short-ranged hopping
Hamiltonian. If the baryons were almost free, we would say that
they are surely doubled like the original quarks. The fact that
the simplicity of the hopping terms is only apparent, and that the
baryons are still coupled strongly to mesonic excitations, offers
the possibility that doubling may not return.

\subsection{Third order: the baryon kinetic term}

The third-order term in $H_{\text{eff}}$, which only exists in the
case of $N_c=3$, is calculated via
\begin{equation}
H_{\text{eff}}^{(3)}=P_0V_FDV_FDV_FP_0.
\end{equation}
For a single link, we have
\begin{equation}
\langle0|U_{\alpha\beta}U_{\gamma\delta}U_{\epsilon\zeta}|0\rangle=
\frac16
\epsilon_{\alpha\gamma\epsilon}\epsilon_{\beta\delta\zeta},
\end{equation}
and the same can be proven for a chain of links,
\begin{equation}
\langle0| \left(\prod U\right)_{\alpha\beta} \left(\prod
U\right)_{\gamma\delta} \left(\prod U\right)_{\epsilon\zeta}
|0\rangle= \frac16
\epsilon_{\alpha\gamma\epsilon}\epsilon_{\beta\delta\zeta}.
\end{equation}
Thus [$f,g,\ldots$ are here (temporarily) flavor indices],
\begin{equation}
H_{\text{eff}}^{(3)}=-i\sum_{j>0}\tilde K(j) \sum_{\bn\mu}
\left(\psi^{\dag f\alpha}_\bn \alpha_\mu
\psi^{f\beta}_{\bn+j\muhat}\right) \left(\psi^{\dag g\gamma}_\bn
\alpha_\mu \psi^{g\delta}_{\bn+j\muhat}\right) \left(\psi^{\dag
h\epsilon}_\bn \alpha_\mu \psi^{h\zeta}_{\bn+j\muhat}\right)
\epsilon_{\alpha\gamma\epsilon}\epsilon_{\beta\delta\zeta}
+\text{\em h.c.}
\end{equation}
The kernel is
\begin{equation}
\tilde K(j)=\frac{(D(j))^3}{6\left(\frac12g^2C_F|j|\right)^2}.
\end{equation}
Again, spin diagonalization simplifies the odd-$j$ terms, but not
the even-$j$ terms. The result is
\begin{equation}
H_{\text{eff}}^{(3)}=H_{\text{odd}}^{(3)}+H_{\text{even}}^{(3)},
\end{equation}
with
\begin{equation}
H_{\text{odd}}^{(3)}= -i\sum_{j>0\atop\text{$j$ odd}} \tilde K(j)
\sum_{\bn\mu}b_\bn^{\dag I}b_{\bn+j\muhat}^I
\eta_\mu(\bn)+\text{\em h.c.},
\end{equation}
where $\eta_\mu(\bn)$ is the usual staggered-fermion sign factor,
and
\begin{equation}
H_{\text{even}}^{(3)}= -i\sum_{j>0\atop\text{$j$ even}} \tilde
K(j) \sum_{\bn\mu}b_\bn^{\dag I}\left[
\alpha_\mu\otimes\alpha_\mu\otimes\alpha_\mu\right]_{II'}
b_{\bn+j\muhat}^{I'}\zeta_\mu(\bn)+\text{\em h.c.},
\end{equation}
where $\zeta_\mu(\bn)=(-1)^{\sum_{\nu\not=\mu}n_\nu}$. The baryon
operators are
\begin{equation}
b^I=\epsilon_{\alpha\beta\gamma}\psi^{\alpha f}\psi^{\beta
g}\psi^{\gamma h},
\end{equation}
where we have written $I$ to represent the compound index
$\{fgh\}$, taking values in the symmetric three-index
representation of $U(N)$ ($f,g,\ldots$ once more combine flavor
and Dirac indices). The odd-$j$ part of $H^{(3)}_{\text{eff}}$,
like that of $H^{(2)}_{\text{eff}}$, is symmetric under the $U(N)$
doubling symmetry. The even-$j$ part breaks $U(N)$ to
$U(N_f)_L\times U(N_f)_R$.

$H^{(3)}_{\text{eff}}$ is a baryon hopping term. As mentioned in
the Introduction, however, its simplicity is deceptive. The baryon
operators $b^I_\bn$ are composite and hence do not obey canonical
anticommutation relations, {\em i.e.,}
\begin{equation}
\left\{b^I_\bn,b^{\dag
I'}_{\bn'}\right\}\not=\delta_{II'}\delta_{\bn\bn'}.
\end{equation}
The separation of $H^{(3)}_{\text{eff}}$ into a canonical kinetic
energy and an interaction term is a challenge for the future.

\section{$\sigma$ model representation \label{sec:NLS}}

Because of the complexity of the third-order effective
Hamiltonian, we restrict ourselves henceforth to the second-order
theory, in which baryons are a fixed background. The theory
defined by $H^{(2)}_{\text{eff}}$ is a generalized spin model,
with spins chosen to be in representations of $U(N)$ according to
the baryon distribution. We review \cite{RS1} in this section how
to convert the spin model into a $\sigma$ model. The $\sigma$
field at each site will move in a manifold determined by the
baryon number at that site.

\subsection{Coherent state basis}

We employ a generalization of spin coherent states \cite{Klauder}
to derive a path integral for the spin model of
$H^{(2)}_{\text{eff}}$. We recall that a given site carries
generators $Q^\eta_\bn$ of $U(N)$ in a representation with $N_c$
columns and $m$ rows, with $B'_\bn=m$.

First we choose a basis for the Lie algebra of $U(N)$. This
consists of the generators $S^i_j$, with $i,j=1,\ldots,N$, whose
matrix elements in the fundamental rep are
\begin{equation}
(S^i_j)_{fg}= \delta_{if}\delta_{jg}.
\end{equation}
The corresponding charges are
\begin{eqnarray}
Q^i_j&=&\sum_\alpha \psi^{\dag}_\alpha S^i_j
\psi_\alpha-\frac12N_c\delta^i_j \nonumber\\
&=&\sum_\alpha \psi^{\dag}_{i \alpha}\psi_{j
\alpha}-\frac12N_c\delta^i_j, \label{Shat}
\end{eqnarray}
where we have subtracted a constant for convenience. The Cartan
subalgebra consists of the operators
\begin{equation}
H_i=Q^i_i. \label{Cartan}
\end{equation}

We build the coherent states from the state of highest weight. The
highest-weight state $|\Psi_0\rangle$ in the representation is an
eigenstate of the Cartan generators,
\begin{equation}
H_i|\Psi_0\rangle=\left\{
\begin{array}{cl}
(N_c/2)|\Psi_0\rangle&{\rm for}\ i=1,\ldots,m\\[2pt]
-(N_c/2)|\Psi_0\rangle&{\rm for}\ i=m+1,\ldots,N.
\end{array}\right.
\label{CartanH}
\end{equation}
In this state, the generators take the simple form
\begin{equation}
\langle \Psi_0|Q^i_j|\Psi_0\rangle=\frac12N_c\Lambda_{ij},
\label{SQ1}
\end{equation}
with
\begin{equation}
\Lambda=\left(
\begin{array}{cc}
\bm1_m&0\\
0&-\bm1_{N-m}
\end{array}\right).
\end{equation}
The state $|\Psi_0\rangle$ is invariant (up to a phase) under the
subgroup of $U(N)$ that commutes with $\Lambda$; this is
$U(m)\times U(N-m)$. The most general rotation of $|\Psi_0\rangle$
is carried out with the generators $Q^i_j$ that are {\em not\/} in
the corresponding subalgebra. We choose parameters
$a_\mu^\lambda$, with $\lambda\in[1,m]$ and $\mu\in[m+1,N]$, and
write
\begin{equation}
|a\rangle=\exp\left(\sum_{\lambda=1}^m \sum_{\mu=m+1}^N
(a_\mu^\lambda Q^\mu_\lambda -a_\mu^{*\lambda}Q_\mu^\lambda)
\right)|\Psi_0\rangle. \label{cohstate}
\end{equation}
The only generators $Q^\mu_\lambda$ that appear in \Eq{cohstate}
are those that lower an $H_i$ that starts from $N_c/2$ in
\Eq{CartanH} while raising another $H_i$ that starts from
$-N_c/2$. Any other generator would annihilate $|\Psi_0\rangle$
and thus give no effect in the exponential.

The coherent states are normalized,
\begin{equation}
\langle a|a\rangle=1,
\end{equation}
and over-complete,
\begin{equation}
\int da\,|a\rangle\langle a|=1. \label{complete}
\end{equation}
In \Eq{complete} the integral is over the coset space
$U(N)/[U(m)\times U(N-m)]$ (see below). Matrix elements of the
generators are given by
\begin{equation}
\langle a|Q^i_j|a\rangle=\frac12N_c\sigma_{ij}, \label{SQ}
\end{equation}
where the matrix $\sigma_{ij}$ is given by a unitary rotation from
$\Lambda$,
\begin{equation}
\sigma=U(a)\Lambda U(a)^{\dag}. \label{QULU}
\end{equation}
The matrix $U(a)$ is built out of the $m\times(N-m)$ matrix
$a_\mu^\lambda$,
\begin{equation}
U=\exp\left[\left(
\begin{array}{cc}
0&a\\
-a^{\dag}&0
\end{array}\right)\right].
\label{Ua}
\end{equation}
$\sigma$ is both Hermitian and unitary.

The manifold of matrices $\sigma$ is the coset space
$U(N)/[U(m)\times U(N-m)]$, a sub-manifold of $U(N)$. This is
because for any matrix $U(a)$, one can generate an orbit $U(a)V$
by multiplying with a matrix
\begin{equation}
V=\left(
\begin{array}{cc}
X&0\\0&Y
\end{array}\right),
\end{equation}
where $X\in U(m)$ and $Y\in U(N-m)$. All matrices in the orbit
will give the same matrix $\sigma$ when inserted into \Eq{QULU},
and thus in integrating over the configuration space of $\sigma$
one must choose only a single representative of each orbit. This
set of representatives, the coset space $U(N)/[U(m)\times
U(N-m)]$, is the quotient space of the non-invariant subgroup
$U(m)\times U(N-m)$.

The measure over the coset space must be invariant under unitary
rotations,
\begin{equation}
|a\rangle\to R(V)|a\rangle,
\end{equation}
where $R(V)$ represents the rotation $V$ in Hilbert space.
Equation~(\ref{SQ}) shows that this is a rotation
\begin{equation}
\sigma\to V\sigma V^{\dag}
\end{equation}
and by \Eq{QULU}, this means that a measure in $U$ must be
invariant under $U\to VU$. This fixes the measure uniquely to be
the Haar measure in $U(N)$, and thus one can integrate over the
coset space by integrating with respect to $U$ over $U(N)$ and
using \Eq{QULU}.

A representation whose Young diagram has $N-m$ rows is the
conjugate to the representation with $m$ rows. Its coherent state
space can be constructed to look the same, with only a sign
difference. To do this we start with the {\em lowest\/}-weight
state, which satisfies [cf.~\Eq{CartanH}]
\begin{equation}
H_i|\Psi_0\rangle=\left\{
\begin{array}{cl}
-(N_c/2)|\Psi_0\rangle&{\rm for}\ i=1,\ldots,m\\[2pt]
(N_c/2)|\Psi_0\rangle&{\rm for}\ i=m+1,\ldots,N.
\end{array}\right.
\end{equation}
This introduces a minus sign into \Eq{SQ1}. The subsequent steps
are identical, with only the replacement of \Eq{SQ} by
\begin{equation}
\langle a|Q^i_j|a\rangle=-\frac12N_c\sigma_{ij}. \label{SQminus}
\end{equation}
Here, too, $\sigma$ is given in terms of $\Lambda$ and $U$ by
\Eq{QULU}.

\subsection{Partition function}

The partition function $Z=\Tr e^{-\beta H}$ can be written as a
path integral by inserting the completeness relation
(\ref{complete}) at every slice of imaginary time. This gives
\begin{equation}
Z=\int D\sigma(\tau)\,\exp -S, \label{partitionfunction}
\end{equation}
where the action is
\begin{equation}
S=\int_0^\beta d\tau\left[\frac{1-\langle
a(\tau)|a(\tau+d\tau)\rangle}{d\tau} +H(\sigma(\tau))\right].
\end{equation}
The Hamiltonian $H(\sigma)$ is a transcription of  the quantum
Hamiltonian to the classical $\sigma$ matrices. Starting with the
quantum operator $Q^\eta_{\bn}$, we have
\begin{eqnarray}
Q^\eta_{\bn}&=&\psi^{\dag}_{\bn}M^\eta\psi_{\bn}
=M^\eta_{ij}\psi^{\dag}_{\bn}S^i_j\psi_{\bn}\nonumber\\
&=&M^\eta_{ij}Q^i_j(\bn)+\frac12N_c\Tr M^\eta.
\end{eqnarray}
Expressed in these variables, the quantum Hamiltonian
is\footnote{The $B'$ term from \Eq{H2eff2} indeed disappears. We
have dropped an additive constant that is independent of $B'$.}
\begin{equation}
H_{\text{eff}}^{(2)}= \sum_{\bn\mu \atop j\not=0}J_j Q^\eta_{\bn}
Q^\eta_{\bn+j\muhat} \left(s^\mu_\eta\right)^{j+1}.
\label{Hquantum}
\end{equation}
where $J_j=(2/N_c)K(j)$. We transcribe this according to \Eq{SQ}
to obtain the classical Hamiltonian,
\begin{equation}
H(\sigma)=\left(\frac{N_c}2\right)^2\sum_{\bn\mu \atop j\not=0}J_j
\sigma^\eta_{\bn}\sigma^\eta_{\bn+j\muhat}\left(s^\mu_\eta\right)^{j+1},
\label{Hsigma}
\end{equation}
where
\begin{equation}
\sigma^\eta_{\bn}=\Tr M^{\eta T} \sigma_\bn.
\end{equation}
Recall that each $\sigma_\bn$ is an $N\times N$ matrix ranging
over the coset space appropriate to site $\bn$.

The time-derivative term in $S$ is a Berry phase \cite{RS1}. It
can be expressed in terms of the matrix $U$ that determines
$\sigma$ via \Eq{QULU}. The result is\footnote{This is correct
only if $U$ is of the form given in \Eq{Ua}, and in that case $U$
cannot be integrated over all of $U(N)$.}
\begin{equation}
S=\int_0^\beta d\tau\left[-\frac{N_c}2\sum_{\bn}\Tr\Lambda_\bn
U^{\dag}_{\bn}
\partial_\tau U_{\bn}
+H(\sigma(\tau))\right]. \label{action_sigma}
\end{equation}
$\Lambda_\bn$ will vary from site to site if $m$ does. If one
takes the route of \Eq{SQminus} for a representation with $N-m$
rows, then the kinetic term for that site acquires a minus sign
(see below).

The number of colors has largely dropped out of the problem, since
$\sigma$ is just an $N\times N$ unitary matrix field. The explicit
factors of $N_c$ in Eqs.~(\ref{Hsigma}) and~(\ref{action_sigma})
invite a semiclassical approximation in the large-$N_c$ limit.
This of course neglects the $N_c$-dependence of the couplings
$J_j\sim 1/N_c^2$, but the common scale of the couplings only
serves to set an energy scale. The ground state will be
independent of this scale, although correlations and the
temperature scale will reflect it. We take the point of view that
after all $N_c=3$, and we are interested in properties of the
effective theory for this value only. The $N_c\to\infty$ limit,
for fixed couplings $J_j$, will be a device for investigating the
properties of a generalized  effective theory.

\section{Zero density \label{sec:zeroB}}

The simplest way to set the baryon density to zero is just to
choose $B_\bn=0$ on each site, meaning $m=N/2=2N_f$. It turns out
to be just as easy to consider a slightly generalized case
\cite{Smit}, in which $B_\bn$ is chosen to alternate, $B_\bn=\pm
b$, on even and odd sublattices. This means to choose a
representation with $m=N/2+b$ rows on even sites and $N-m=N/2-b$
rows on odd sites, which gives a pair of conjugate representations
of $U(N)$. In view of \Eq{SQminus}, we can  substitute
$\sigma\to-\sigma$ on the odd sites and thus have identical
manifolds on all sites. The Hamiltonian is then
\begin{equation}
H(\sigma)=\left(\frac{N_c}2\right)^2\sum_{\bn\mu \atop
j\not=0}J_j(-1)^j
\sigma^\eta_{\bn}\sigma^\eta_{\bn+j\muhat}\left(s^\mu_\eta\right)^{j+1}.
\label{Hsigmaminus}
\end{equation}
$m$ is a new parameter in the theory, and we can ask what value of
$m$ gives the lowest energy for the ground state. We will see that
$m=N/2$ is indeed preferred, but only in the next-to-leading order
in $1/N_c$.

\subsection{Large-$N_c$ limit}

In the large-$N_c$ limit, we seek the classical saddle
point\footnote{Note that the kinetic term is pure imaginary.} of
$S$. We assume the saddle is at a configuration
$\sigma_{\bn}(\tau)$ that is independent of time, and so we drop
the time derivative. We begin with the nearest-neighbor
Hamiltonian,\footnote{ This classical Hamiltonian is
ferromagnetic; the antiferromagnetic nature of the quantum
Hamiltonian (\ref{Hquantum}) is preserved by the alternating signs
in the time-derivative term.}
\begin{eqnarray}
H&=&-J\sum \sigma^\eta_{\bn}\sigma^\eta_{\bn+\muhat}\nonumber\\
&=&-\frac J2\sum \Tr \sigma_\bn \sigma_{\bn+\muhat}.
\end{eqnarray}
Again, the matrices $\sigma_\bn$ are Hermitian and unitary, and
the expansion coefficients $\sigma^\eta_{\bn}$ are real and
satisfy $\sum(\sigma^\eta)^2=N/2$. The minimum of $H$ is clearly
at a constant field, $\sigma_\bn=\sigma_0$, which can be
diagonalized to $\sigma_\bn=\Lambda$ by a $U(N)$ rotation. This is
a N\'eel state in the original variables. The $U(N)$ symmetry is
broken to $U(m)\times U(N-m)$ and there are $2m(N-m)$ Goldstone
bosons.

The classical energy density (per link) is $\epsilon_0=-JN/2$,
independent of $m$. Thus at leading order in $1/N_c$, the optimal
value of $m$ is undetermined, and any alternating background of
baryon number is equally good.

Since it turns out that the $1/N_c$ corrections select $m=N/2$,
let us consider the effect of the \nnn\ term in $H$ for this case
only. The perturbation is
\begin{equation}
H'=J'\sum_{\bn\mu}
\sigma^\eta_{\bn}\sigma^\eta_{\bn+\muhat}s^\mu_\eta.
\end{equation}
It breaks the $U(N)$ symmetry to $SU(N_f)_L\times SU(N_f)_R\times
U(1)_A\times U(1)_B$. Assuming that $J'\ll J$, we again seek the
minimum energy configuration in the form of a constant field; we
minimize
\begin{equation}
\epsilon'=\sum_{\mu\eta} \sigma^\eta \sigma^\eta s^\mu_\eta
\label{epsprime}
\end{equation}
among the $U(N)$-equivalent $\sigma=\sigma_0$ states that minimize
the nearest-neighbor action. It is not hard to show (see Appendix
\ref{app:nnn}) that $\epsilon'$ is minimized for
$\sigma=\gamma_0\otimes\bm{1}$. This is a condensate that is
symmetric under the vector generators
$M^\eta=\bm{1}\otimes\lambda^a$ but not under the axial generators
$\gamma_5\otimes\lambda^a$, and thus it breaks the chiral symmetry
to the vector subgroup, $SU(N_f)_V\times U(1)_B$.

\subsection{$1/N_c$ corrections}

Returning to the nearest-neighbor theory, we consider fluctuations
around the $\sigma_\bn=\Lambda$ minimum of $S$. First we rescale
$\tau\to2\tau/N_c$ in order to put the kinetic and potential terms
on an equal footing, giving
\begin{equation}
S=\frac{N_c}2\int_0^{\bar\beta} d\tau\left[-\sum_{\bn}(-1)^{\bn}
\Tr\Lambda U^{\dag}_{\bn} \partial_\tau U_{\bn} -\frac
J2\sum_{\bn\mu} \Tr \sigma_\bn \sigma_{\bn+\muhat}\right],
\label{Srescaled}
\end{equation}
with $\bar\beta=N_c\beta/2$. Recalling \Eq{Ua}, we can write
\begin{equation}
U_\bn=e^{A_\bn}, \label{expand1}
\end{equation}
where $A_\bn$ is anti-Hermitian and anticommutes with $\Lambda$.
It is more convenient to work with the Hermitian matrix
\begin{equation}
L_\bn=2A_\bn\Lambda,
\end{equation}
in terms of which we expand
\begin{equation}
\sigma_\bn=\Lambda+L_\bn-\frac12L^2\Lambda. \label{sigmaL}
\end{equation}
If we further expand $L_\bn$ in the basis of generators of $U(N)$,
\begin{equation}
L_\bn=\sum_\eta l^\eta M^\eta, \label{Ll_exp}
\end{equation}
we find that the $l^\eta$ corresponding to generators of
$U(m)\times U(N-m)$ vanish; this is the subgroup under which the
vacuum is symmetric. The field $L_\bn$ thus contains $2m(N-m)$
real degrees of freedom, corresponding to the Goldstone bosons.

We expand $U_\bn$ and $\sigma_\bn$ in powers of $L_\bn$; using
\Eq{Ll_exp}, we obtain to second order
\begin{equation}
S=S_0+\frac{N_c}2\int d\tau \sum_\bn\left[
\frac{(-1)^\bn}8C^{\eta\eta'}l_\bn^\eta \partial_\tau
l_\bn^{\eta'} +\frac J8\sum_\mu\left(l_{\bn+\muhat}-l_\bn
\right)^2\right].
\end{equation}
The coefficient matrix is
\begin{equation}
C^{\eta\eta'}=\Tr(\Lambda[M^\eta,M^{\eta'}]), \label{Cetaeta}
\end{equation}
and the classical energy is $S_0=-\frac
J2\left(\frac{N_c}2\right)^2NN_sd\beta$. $C$ is antisymmetric and
purely imaginary; we show in Appendix \ref{app:CD} that $C$ has
eigenvalues $\pm1$, each with degeneracy $m(N-m)$. We change basis
so as to diagonalize $C$, and write the index $\eta$ as the
compound $(\alpha,\pm)$ with $\alpha=1,\ldots,m(N-m)$ and the
$\pm$ corresponding to the eigenvalue of $C$. Since the original
$l^\eta$ are real, we have
\begin{equation}
(l^{\alpha+})^*=l^{\alpha-}.
\end{equation}
Thus we eliminate $l^{\alpha-}$ and write
\begin{equation}
S=S_0+\frac{N_c}8\int d\tau \sum_\bn\left[ (-1)^\bn i\,\Im
l_\bn^{\alpha+*}\partial_\tau l_\bn^{\alpha+} +\frac
J2\sum_\mu\left|l_{\bn+\muhat}^{\alpha+}-l_\bn^{\alpha+}\right|^2\right].
\label{Sln}
\end{equation}

The alternating sign in \Eq{Sln} is what makes the theory
antiferromagnetic. It forces us to differentiate between even and
odd sites, and we transform to momentum space as follows (dropping
the $+$ superscript):
\begin{equation}
l^\alpha_\bn=\sqrt{\frac2{N_s}}\sum_\bk\left\{
\begin{array}{ll}
l^\alpha_{1,\bk}e^{i\bk\cdot\bn}&\quad \text{$\bn$ even,}\\[2pt]
l^\alpha_{2,\bk}e^{i\bk\cdot(\bn-\hat z)}&\quad \text{$\bn$ odd.}
\end{array}\right.
\end{equation}
The even sites comprise an {\em fcc\/} lattice with lattice
constant 2, and the momenta $\bk$ take values in its Brillouin
zone. We obtain
\begin{equation}
S=S_0+\frac{N_c}8\sum_\bk\int d\tau\, (l_1^*\quad
l_2^*)_{\bk}^\alpha \,\mathcal{M}(\bk) \left(\begin{array}{c}
l_1\\l_2
\end{array}\right)_{\bk}^\alpha.
\end{equation}
Here
\begin{equation}
\mathcal{M}(\bk)=\left({\renewcommand{\arraystretch}{1.5}
\begin{array}{cc}
Jd-\partial_\tau&-Jd\gamma(\bk)\\[2pt]
-Jd\gamma(\bk)&Jd+\partial_\tau
\end{array}}\right)
\end{equation}
and  $\gamma(\bk)=\frac1d\sum_\mu \cos k_\mu$.

The gaussian path integral over the $l$ field now gives the free
energy,
\begin{equation}
F = F_0+ m(N-m)\frac{N_c}2\sum_\bk \left[ \frac1{\bar\beta}\log
\left( 2\sinh \frac{\bar\beta\omega(\bk)}2\right) - \frac{Jd}2
\right],
\end{equation}
where $\omega(\bk) = dJN_c\sqrt{1-\gamma^2(\bk)}$ and $F_0 =
-\frac12JdNN_s(N_c/2)^2$. For the ground state energy, we take
$\beta\to\infty$ and obtain (restoring all constants)
\begin{equation}
E_0 = -JN_sNd\left(\frac{N_c}{2} \right)^2 \left[ 1 +
\frac{1}{N_c} \frac{m(N-m) }{N} \int_{\rm BZ} \left(
\frac{dk}{2\pi} \right)^d \left( 1-\sqrt{1-\gamma^2(\bk) } \right)
\right]. \label{eq:E0}
\end{equation}
This is exactly the result of Smit \cite{Smit}. The $O(1/N_c)$
corrections lift the degeneracy of the ground states with
different values of $m$. The integral in \Eq{eq:E0} is positive
and its coefficient contains the number of Goldstone bosons. Thus
the state of lowest energy is that with $m=N/2$, and the symmetry
breaking scheme is $U(N)\to U(N/2)\times U(N/2)$. Further breaking
by the \nnn\ terms was discussed above.

\section{Nonzero baryon density\label{sec:nonzeroB}}

The zero-density theories considered in the preceding section were
defined by selecting representations with $m$ and $N-m$ rows on
alternating sites. For any $m$, this led to a $\sigma$ model with
identical degrees of freedom on all sites---after redefinition of
the spins on the odd sublattice---and ferromagnetic
couplings.\footnote{The $(-1)^\bn$ factor in the kinetic energy
retained information about the antiferromagnetic nature of the
quantum problem; it did not affect the classical analysis.} We
eventually settled on $m=N/2$ as the background that gives the
ground state of lowest energy.

Introducing non-zero baryon density means changing $m$ on some
sites of the lattice. Since in general there will be
representations on different sites that are not mutually
conjugate, different sites will carry $\sigma$ variables that do
not live in the same submanifold of $U(N)$. We here limit
ourselves to the simpler case of uniform $m$, where adjacent sites
carry identical spins---but the coupling is {\em
antiferromagnetic}.

In order to learn how to work with such a theory, we begin by
studying the two-site problem. The results of this study will lead
directly to an {\em ansatz\/} for the ground state of a lattice
with a fixed density of baryons.

\subsection{The two-site problem}

\subsubsection{Classical solution}

\begin{figure}
\epsfig{width=10cm,file=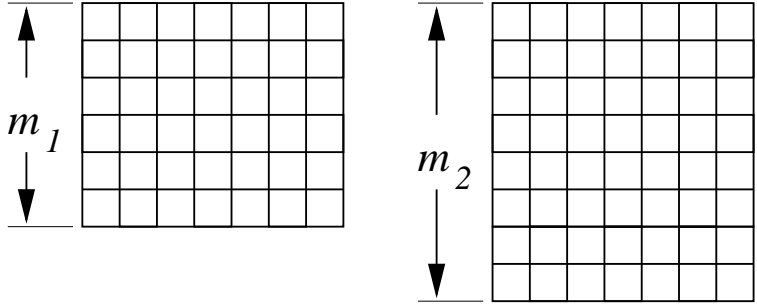} \caption{Two $U(N)$ spins in
different representations} \label{fig:twospins}
\end{figure}

Consider, therefore, two sites with quantum spins $Q_1$ and~$Q_2$
that carry representations of $U(N)$ with $m_1$ and~$m_2$ rows,
and $N_c$ columns (see Fig.~\ref{fig:twospins}).
The quantum Hamiltonian is
\begin{equation}
H=JQ_1^\eta Q_2^\eta, \label{2site_qH}
\end{equation}
an antiferromagnetic coupling. The corresponding classical
$\sigma$ model has the interaction Hamiltonian
\begin{equation}
H(\sigma)=\frac J2 \left(\frac{N_c}2\right)^2 \Tr
\sigma_1\sigma_2,
\end{equation}
where
\begin{equation}
\sigma_i=U_i\Lambda_iU_i^\dag.
\end{equation}
The two $\Lambda$ matrices reflect the different values of $m_i$
according to
\begin{equation}
\Lambda_i=\left(
\begin{array}{cc}
\bm1_{m_i}&0\\
0&-\bm1_{N-m_i}
\end{array}\right).
\end{equation}

The $N_c\to\infty$ limit is the classical limit, in which we seek
values of $\sigma_{1,2}$ that minimize $H(\sigma)$. A global
$U(N)$ rotation, {\em viz.},
\begin{equation}
\sigma_i\to V\sigma_i V^\dag,
\end{equation}
can be used to diagonalize $\sigma_1$ so that
$\sigma_1=\Lambda_1$. Now we have to minimize $\Tr
\Lambda_1\sigma_2$. The case of conjugate representations,
$m_2=N-m_1$, is easy: $\sigma_2$ is a unitary rotation of
$\Lambda_2$, which (in this case) can be rotated into
$-\Lambda_1$. This is the unique antiferromagnetic ground state.
$\sigma_1$ and~$\sigma_2$ can be copied to the odd and even
sublattices of an infinite lattice to give the classical N\'eel
state considered in the preceding section.

The case $m_1=m_2=m$ is more complex. We consider $m>N/2$ for
definiteness; the other case is similarly handled. We write
explicitly\footnote{This generalizes a parametrization found in
\cite{SW}.} [from Eqs.~(\ref{QULU})--(\ref{Ua})]
\begin{equation}
\sigma_2=\left( \begin{array}{cc} \cos\left(2\sqrt{a
a^\dag}\right) & -a\frac{\displaystyle\sin\left(2\sqrt{a^\dag
a}\right)}
      {\displaystyle\sqrt{a^\dag a}}       \\[2pt]
-\frac{\displaystyle\sin\left(2\sqrt{a^\dag a}\right)}
            {\displaystyle\sqrt{a a^\dag}}\,a^\dag &
-\cos\left(2\sqrt{a^\dag a}\right)
\end{array}\right).
\label{eq:general_Q}
\end{equation}
$a^{\dag} a$ is a square matrix of dimension $N-m$ and $a
a^{\dag}$ is a square matrix of dimension $m$. Since $\sigma_2$ is
a rotation of $\Lambda$,
\begin{equation}
2m-N=\Tr\sigma_2= \Tr\cos\left(2\sqrt{a
a^\dag}\right)-\Tr\cos\left(2\sqrt{a^\dag a}\right),
\label{eq:B_restrict}
\end{equation}
and hence the energy is
\begin{eqnarray}
E&=&\frac{J}2 \left(\frac{N_c}2\right)^2\Tr\Lambda\sigma\\
&=&J\left(\frac{N_c}2\right)^2\left[\Tr\cos\left(2\sqrt{a^\dag
a}\right) +2m-N\right].
\end{eqnarray}
$E$ is minimized when all the eigenvalues of $a^\dag a$ are equal
to $\pi^2/4$. This means that the $N-m$ column vectors $\ba_i$
form an orthogonal set in $m$ dimensions, with
\begin{equation}
\ba_i^\dag\cdot\ba_j=\left(\frac\pi2\right)^2\delta_{ij}.
\end{equation}
Since $m>N-m$ by assumption, such a set of vectors can always be
found.

Since $a^\dag a=(\pi^2/4)\bm1_{N-m}$, we have
$\sin\left(2\sqrt{a^\dag a}\right)=0$ and so the off-diagonal
blocks of \Eq{eq:general_Q} vanish. The lower-right block of
$\sigma_2$ is the unit matrix $\bm1_{N-m}$. We know that
$(\sigma_2)^2=1$ since $\Lambda^2=1$ and thus the upper-left block
must have eigenvalues $\pm1$. Equating traces of $\sigma_2$ and
$\Lambda$, we find that the upper-left $m\times m$ block must take
the form
\begin{equation}
\sigma^{(m)}=U^{(m)}\Lambda^{(m)}U^{(m)\dag},
\end{equation}
where
\begin{equation}
\Lambda^{(m)}=\left( \begin{array}{cc}
\bm1_{2m-N}&0\\
0&-\bm1_{N-m}
\end{array}\right)
\label{eq:Lm}
\end{equation}
and $U^{(m)}\in U(m)$. $\sigma^{(m)}$ represents the coset
$U(m)/[U(2m-N)\times U(N-m)]$.

We conclude that the classical ground state of this $B\not=0$
two-site problem is degenerate, even beyond breaking the overall
$U(N)$ symmetry. The solutions can be written as
\begin{eqnarray}
\sigma_1&=&\Lambda_1=\left( \begin{array}{cc}
\bm1_m&0\\
0&-\bm1_{N-m}\end{array} \right), \nonumber\\[2pt]
\sigma_2&=&\left( \begin{array}{cc}
\sigma^{(m)}     &       0       \\
0       &       \bm1_{N-m} \end{array} \right) \label{eq:Q2}
\end{eqnarray}
(to which a global $U(N)$ rotation can be applied). A particular
instance of $\sigma^{(m)}$ is $\Lambda^{(m)}$, given by
\Eq{eq:Lm}. The symmetry of the vacuum is the set of rotations
that leaves both $\sigma_1$ and $\sigma_2$ invariant, namely,
$U(2m-N)\times U(N-m)\times U(N-m)$.

\subsubsection{Quantum fluctuations}

The classical solution of the two-site problem will guide us in
approaching the problem of an infinite lattice below. We expect
that spontaneous symmetry breaking will give a vacuum of the same
character, with continuous degeneracy. There are, however, two
kinds of degeneracy in the two-site problem: that which results
from breaking the global $U(N)$ to $U(m)\times U(N-m)$, and that
which comes of breaking the $U(m)$ subgroup to $U(2m-N)\times
U(N-m).$ The latter degeneracy is connected with freedom in
choosing the orientation of $\sigma_2$ {\em relative to\/}
$\sigma_1$. It is instructive to see how quantum mechanical
fluctuations lift the classical degeneracies.

\begin{figure}
\epsfig{width=15cm,file=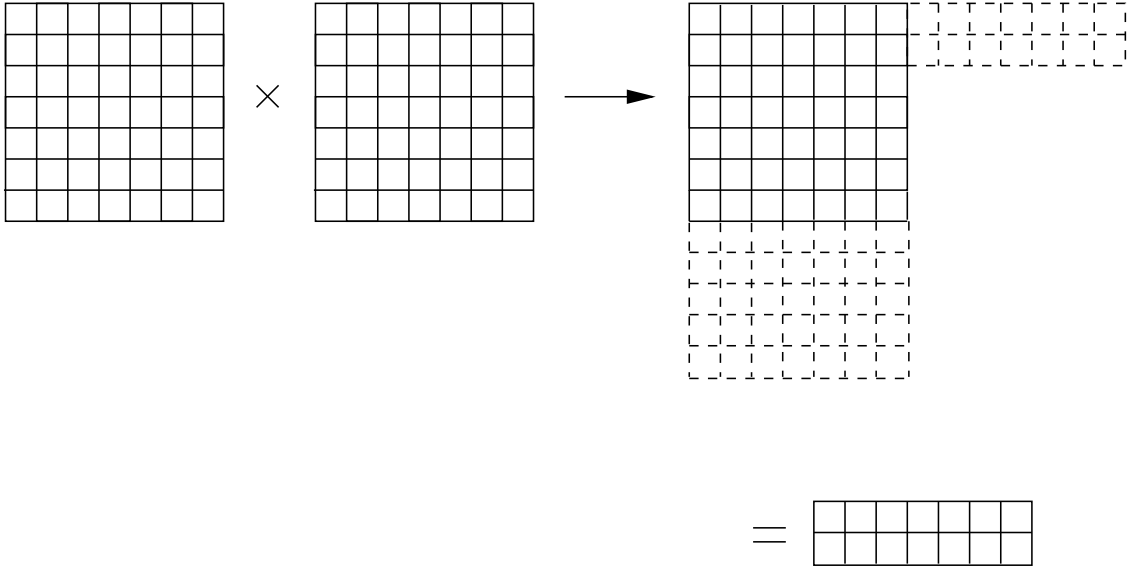} \caption{Coupling two spins
in a 7-row representation of $U(12)$ to the representation with
minimal Casimir operator} \label{fig:twosite}
\end{figure}

The quantum two-site problem is easy to solve. We rewrite the
Hamiltonian (\ref{2site_qH}) as
\begin{equation}
H=\frac J2\left[(Q_1+Q_2)^2-Q_1^2-Q_2^2\right]. \label{2site_qH2}
\end{equation}
$Q_1^2$ and $Q_2^2$ are constants, the quadratic Casimir operator
in the $m$-row, $N_c$-column representation of $U(N)$. The first
term in \Eq{2site_qH2} is minimized by coupling $Q_1$ and $Q_2$ to
the representation that minimizes the Casimir, which is the
representation with $2m-N$ rows and $N_c$ columns (see
Fig.~\ref{fig:twosite}).
The ground state has discrete degeneracy equal to the dimension of
this representation.

The exact quantum solution naturally shows no sign of spontaneous
symmetry breaking and hence it is not of much relevance to the
infinite volume problem. More interesting is the problem where the
state of $Q_1$ is {\em fixed\/} and $Q_2$ is allowed to vary. This
breaks by hand the global $U(N)$ while allowing quantum
fluctuations to lift any remaining degeneracy in the relative
orientation of the two spins, so it can be regarded as
quantization in the presence of spontaneous symmetry breaking. In
effect, this is mean field theory.

\begin{figure}[htb]
\epsfig{file=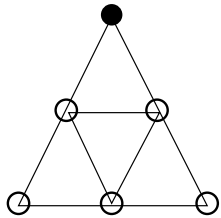} \caption{Weight diagram for $N=3$,
$m=1$, $N_c=2$ [the sextet of $SU(3)$]. The highest-weight state
lies at the top of the triangle; all states at the base minimize
$H^{MF}$. All values of $N_c$ give the same triangular shape, with
multiplicity of one along the boundary. There are $N_c+1$ states
at the base. For $N=4$ the triangle becomes a tetrahedron, and
there are $(N_c+1)(N_c+2)/2$ states at its base.}
\label{fig:twositeweights}
\end{figure}

We replace the Hamiltonian (\ref{2site_qH}) by
\begin{equation}
H^{MF}=J\sum_{\eta=1}^{N^2}\langle Q_1^\eta\rangle Q_2^\eta.
\end{equation}
To minimize the energy we maximize the mean field by choosing the
state of $Q_1$ to be the highest-weight state. This state
diagonalizes the generators $H_i$ of the Cartan subalgebra while
other generators of $U(N)$ have expectation value zero. Thus
\begin{equation}
H^{MF}=J\sum_{i=1}^N\langle H_{1i}\rangle H_{2i}.
\end{equation}
The operators $H_i$ all commute, and their eigenvalues make up the
weight diagram of the representation.\footnote{More precisely, the
weight diagram shows eigenvalues of the $N-1$ traceless diagonal
generators of $SU(N)$.  These can be obtained by isolating the
$U(1)$ member of the set $H_i$ and taking linear combinations of
the rest.} $H^{MF}$ is a dot product of the weight vectors of the
two spins. The energy is minimized by choosing for $Q_2$ a state
that lies opposite the highest-weight state in the weight diagram.
As shown in the example of Fig.~\ref{fig:twositeweights}, this
still leaves a degeneracy, albeit a discrete one. We stress that
this degeneracy comes from freedom in the relative orientation of
$Q_1$ and~$Q_2$; it remains after quantum fluctuations lift the
continuous degeneracy of the classical system.

In the $N_c\to\infty$ limit, the discrete degeneracy becomes
infinite and presumably it is well described by the continuous
degeneracy of the classical problem.

\subsection{Infinite lattice}

At $N_c=\infty$ we seek the saddle point of the action, which we
assume to be a time-independent configuration. The classical
Hamiltonian of the $\sigma$ model is
\begin{equation}
H=\frac J2\sum_{\bn,\mu} \Tr \sigma_\bn \sigma_{\bn+\muhat}.
\end{equation}
Seeking an antiferromagnetic ground state, we set
$\sigma_\bn=\Lambda$ on the sublattice of even sites. The odd
sites are then governed by
\begin{equation}
H^{\text{odd}}=Jd\sum_{\bn\ \text{odd}}\Tr\Lambda\sigma_\bn.
\label{Hodd}
\end{equation}
This is just the two-site problem studied above, replicated over
the lattice. As we saw above, the ground state configuration is
degenerate with respect to the configuration at each odd site,
\begin{equation}
\sigma_\bn=\left( \begin{array}{cc}
\sigma_\bn^{(m)}     &       0       \\
0       &       \bm1_{N-m} \end{array} \right) \label{submanifold}
\end{equation}
A uniform choice for the odd sites,
$\sigma_\bn^{(m)}=\Lambda^{(m)}$ for instance, breaks the $U(N)$
symmetry to $U(2m-N)\times U(N-m)\times U(N-m)$; a non-uniform
choice can break the symmetry all the way to $U(N-m)$. The entropy
of this classical ground state is evidently proportional to the
volume.

As noted for the two-site problem, the continuous degeneracy of
the ground state is an artifact of the classical, $N_c\to\infty$
limit. Quantum fluctuations will spread each odd spin's wave
function over the $U(m)/[U(2m-N)\times U(N-m)]$ manifold. A
mean-field {\em ansatz\/} for the even spins, as noted in the
discussion of the two-spin problem, will still leave a discrete
degeneracy for the odd spins; the symmetry breaking scheme will
depend on how the odd spins are allocated to the available states.
Furthermore, one can contemplate making a non-uniform {\em
ansatz\/} for the {\em even\/} spins as well, reducing the phase
space available for odd spins with unequal neighbors but adding
entropy on the even sublattice. The situation is reminiscent of
that in the antiferromagnetic Potts model \cite{AFPotts}, the
phase structure of which is not yet understood.

This Potts-like discrete degeneracy, however, is an artifact of
the mean-field approach that, like the classical {\em ansatz\/},
assumes a fixed state for the spins on the even sites. An
essential difference between our $\sigma$ model and the Potts
model is that our degrees of freedom are continuous and will
fluctuate as soon as they are allowed to do so. A given odd spin
will {\em not\/} be surrounded by a uniform fixed background; the
neighboring even spins will be influenced by {\em all\/} their odd
neighbors, and will induce an interaction among the odd spins that
makes them rotate together. This should reduce the entropy of the
ground state to zero. The systematic way to see this effect is to
carry out a $1/N_c$ expansion around the classical {\em ansatz\/},
which we do in Appendix \ref{app:corrections}. The result is a
ferromagnetic interaction among the $\sigma_\bn$ on the odd sites.
Thus the ground state turns out to be the two-site solution,
replicated {\em uniformly\/} over the lattice:
\begin{equation}
\begin{array}{rcll}
\sigma_\bn&=&\Lambda,\hfill&\ \text{$\bn$ even}, \\[2pt]
\sigma_\bn&=&\left(\begin{array}{cc}
\sigma^{(m)}&0\\
0&\bm1_{N-m}
\end{array}\right),&\ \text{$\bn$ odd},
\end{array}
\end{equation}
where $\sigma^{(m)}=U^{(m)}\Lambda^{(m)}U^{(m)\dag}$ is a {\em
global\/} degree of freedom. The symmetry group of the vacuum is
$U(2m-N)\times U(N-m)\times U(N-m)$.

\section{Summary and discussion \label{sec:summary}}

Let us summarize the results presented in this paper.
In the vacuum sector, we have rederived Smit's result for the
lowest-energy configuration of alternating $B_\bn=\pm(m-N/2)$ sites.
The result is indeed $B_\bn=0$; the $U(4N_f)$ symmetry of the nearest-neighbor
theory is spontaneously broken to $U(2N_f)\times U(2N_f)$.
We extended this result to the \nnn\ theory and found that its
$U(N_f)\times U(N_f)$ chiral symmetry is broken to the vector
$U(N_f)$ flavor subgroup.
Adding net baryon number to the system, we examined the case of
uniform baryon density, $B_\bn=m-2N_f>0$.
Here our study was limited to the nearest-neighbor theory, and we
found a N\'eel-like ground state that breaks
$U(N)$ to $U(2m-N)\times U(N-m) \times U(N-m)$.
The number of Goldstone bosons $n_{GB}$ thus depends on the baryon
density $B_\bn$ as 
\begin{equation}
n_{GB}=2(2N_f-|B_\bn|)(2N_f+3|B_\bn|).
\end{equation}

Directions for future work begin with adding \nnn\ interactions to
the $B>0$ theory and extracting from it a prediction for the
breaking of the continuum-like chiral symmetry.
(The axial $U(1)$ symmetry can be broken by hand.)
Another direction is to gain greater freedom in fixing the
baryon density.
A constant value of $B_\bn>0$ means a baryon density that is close
to the maximum allowed on the lattice; the density can be lowered by
setting $B_\bn\not=0$ only on a sparse sublattice, along the lines
shown in \cite{Boston}.
An ultimate goal, as for the Hubbard model, is the incorporation of
the third-order term in the effective Hamiltonian in order to have
a theory with dynamical baryons.
Perhaps an instructive half-measure would be to study the second-order
theory in the presence of a disordered baryon background.

The strong-coupling effective theory can be regarded as a QCD-like model,
possessing gauge invariance and the correct degrees of freedom.
In that case the lattice spacing is merely a parameter, an overall scale.
More insight can be gained by considering the strong-coupling
theory to represent QCD at large distances, derived by some
renormalization-group transformation from a weak-coupling short-distance
Hamiltonian.
On the one hand, one would expect any such effective Hamiltonian to contain
many terms of great complexity; on the other hand, a simple lattice
theory such as ours might offer a qualitative approximation to the
real theory (as long as one accepts the loss of Lorentz invariance).
We can estimate the lattice spacing to be some scale
at which the running QCD coupling is large, certainly greater than the
radius of a proton.
The limitation that the lattice puts on the density then becomes a physical
issue.
Taking the lattice spacing $a$ to be on the order of 1~fm, the highest
baryon density allowed by the lattice is $2N_f$~fm$^{-3}$.
For low values of $N_f$ this may not be enough to see finite-density
phase transitions, in particular a transition to color superconductivity.
Perhaps a way out is to consider an unphysically large number of flavors.

\begin{acknowledgments}
This work was begun while B.S. was on sabbatical leave at the
Center for Theoretical Physics at MIT.  He thanks the members of
the center for their hospitality.  B.B. thanks the organizers of
the 19th Jerusalem Winter School in Theoretical Physics for their
hospitality.  We thank Assa Auerbach, Richard Brower, Krishna
Rajagopal, Subir Sachdev, and Uwe-Jens Wiese for helpful
discussions. This work was supported by the U.S. Department of
Energy under grant no.~DF-FC02-94ER40818 and by the Israel Science
Foundation under grant no.~222/02-1.
\end{acknowledgments}

\appendix
\section{Minimizing the next-nearest-neighbor term \label{app:nnn}}
\label{sec:App1}

The $s^\mu_\eta$ signs are defined only when the $\sigma^\eta$ are
written in the basis
$M^\eta=\Gamma^A\otimes\lambda^a=\rho^\alpha\otimes\sigma^\beta
\otimes\lambda^a$. We choose a chiral basis for the gamma
matrices, so that $\gamma_5=\rho^3$, $\alpha_i=\rho^3\sigma^i$,
and $\beta=\rho^1$. The energy (\ref{epsprime}) is a sum of
squares,
\begin{equation}
\epsilon'=\sum_{\eta} A_\eta \left(\sigma^\eta\right)^2,
\end{equation}
with the constraint $\sum_\eta (\sigma^\eta)^2=N/2$. The
coefficients $A_\eta=\sum_\mu s^\mu_\eta$ take on the values
$\{-3,-1,1,3\}$. The minimum of $\epsilon'$ occurs when all
$\sigma^\eta$ are zero except those corresponding to $A_\eta=-3$,
namely, those for
$M^\eta=\beta\otimes\lambda^a=\rho^1\otimes\lambda^a$ and
$M^\eta=\beta\gamma_5\otimes\lambda^a=\rho^2\otimes\lambda^a$; the
energy is independent of these $\sigma^\eta$. Thus the set of
solutions can be written in the form
\begin{equation}
\sigma_0=\left(
\begin{array}{cccc}
0&0&U&0\\
0&0&0&U\\
U^{\dag}&0&0&0\\
0&U^{\dag}&0&0
\end{array}
\right) =\frac{\rho^1+i\rho^2}2\otimes U
+\frac{\rho^1-i\rho^2}2\otimes U^{\dag}.
\end{equation}
Recalling that $\sigma_0^2=1$, we have $UU^\dag=\bm1$, so $U\in
U(N_f)$. A chiral rotation $\sigma_0\to V^{\dag}\sigma_0V$, with
\begin{equation}
V= \left( \begin{array}{cccc}
U&0&0&0\\
0&U&0&0\\
0&0&{\bf 1}&0\\
0&0&0&{\bf 1}
\end{array}\right)=
\frac12(1+\rho^3)\otimes U+\frac12(1+\rho^3)\otimes{\bf 1},
\end{equation}
turns $\sigma_0$ into $\rho^1=\gamma_0$, which is invariant only
under vector transformations generated by $\bm1\otimes\lambda^a$.

\section{The matrices $C$ and $D_\bn$ \label{app:CD}}
\label{sec:App2}

To analyze the matrices $C$ and $D_\bn$, given by
Eqs.~(\ref{Cetaeta}) and~(\ref{Dmatrix}), we begin with the $U(N)$
generators $M^\eta$ that lie outside the subalgebra $H=U(m)\times
U(N-m)$ that commutes with
\begin{equation}
\Lambda=\left(\begin{array}{cc}
\bm1_m&0\\
0&-\bm1_{N-m}
\end{array}\right).
\end{equation}
We choose for them a basis $M^{pq1}$ and $M^{pq2}$ with
$p=1,\dots,m$ and $q=m+1,\dots,N$, given by
\begin{eqnarray}
(M^{pq1})_{fg} &=& \frac 12\left(\delta_{pf}\delta_{qg}
+\delta_{p g}\delta_{qf}\right) \label{eq:M1} \\
(M^{pq2})_{fg} &=& \frac i2\left(\delta_{pf}\delta_{qg} -\delta_{p
g}\delta_{qf}\right). \label{eq:M2}
\end{eqnarray}
Since the coset space $U(N)/H$ is a symmetric space, the
commutator $[M^{pqa},M^{p'q'a'}]$ lies in $H$; in order for
$\Tr(\Lambda[M^{pqa},M^{p'q'a'}])$ to be nonzero, the commutator
must have a nonzero component in the Cartan subalgebra of $H$.
This is only possible if $a\not=a'$ and $(p,q)=(p',q')$. Thus in
this basis $C$ takes the form
\begin{equation}
C=i\left(\begin{array}{cc}
0&\bm1_{m(N-m)}\\
-\bm1_{m(N-m)}&0
\end{array}\right).
\end{equation}
Diagonalizing $C$ gives eigenvalues $\pm1$. The generators
corresponding to the basis that diagonalizes $C$ are
\begin{eqnarray}
(M^{pq+})_{f g} &=& (M^{pq1} + i M^{pq2})_{f g} = \delta_{pg}\delta_{qf}     \\
(M^{pq-})_{f g} &=& (M^{pq1} - i M^{pq2})_{f g} =
\delta_{pf}\delta_{qg}.
\end{eqnarray}

As for $D$: The only anticommutators among the $M^{pq\pm}$ that do
not vanish [note the bounds on $(p,q)$] are between $M^{pq+}$ and
$M^{p'q'-}$, {\em viz.,}
\begin{equation}
\{M^{pq+},M^{p'q'-}\}_{fg} =\delta_{qq'}\delta_{pf}\delta_{p'g} +
\delta_{pp'}\delta_{q'f}\delta_{qg}. \label{eq:antcomm}
\end{equation}
Noting that $M^{pq-}=(M^{pq+})^\dag$, we find that $D_\bn$ takes
the block-diagonal form
\begin{eqnarray}
D_\bn^{pq\pm,p'q'\pm}&=& -\sum_{\bmm(\bn)}\Tr \left[\{
(M^{pq\pm})^\dag,M^{p'q'\pm}\} \left( \begin{array}{cc}
        \sigma^{(m)}_\bmm      &       0       \\
        0       &       - \bm1_{N-m}  \end{array} \right) \right]     \\
&=&\delta_{qq'} \sum_{\bmm(\bn)}
(\bm1_{m}-\sigma^{(m)}_\bmm)_{pp'}.
\end{eqnarray}
We summarize this by writing
\begin{equation}
D =  \left( \begin{array}{cc}
        E_\bn &  0       \\
        0       &       E_\bn \end{array} \right) \otimes 1_{N-m},
\end{equation}
where $E_\bn$ is an $m\times m$ matrix given by
\begin{equation}
E_\bn = \sum_{\bmm(\bn)} \left( \bm1_m - \sigma^{(m)}_\bmm
\right). \label{eq:E}
\end{equation}
It is easy to prove that the eigenvalues of $E$ range from $0$ to
$4d$. In particular, $D$ is positive.

\section{$1/N_c$ corrections to the $B\not=0$ problem \label{app:corrections}}
\label{sec:App3}

We build on the $N_c=\infty$ vacua described in
Sec.~\ref{sec:nonzeroB} by allowing fluctuations around them. We
let the $\sigma_\bn$ on the even sites fluctuate around $\Lambda$;
we let the $\sigma_\bn$ on odd sites roll freely around the
$U(m)/[U(2m-N)\times U(N-m)]$ manifold covered by
\Eq{submanifold}, and also execute small oscillations off the
manifold into the $U(N)/[U(m)\times U(N-m)]$ coset space. Our goal
is an effective action for the classical part (\ref{submanifold})
of the odd spins. To reach this, we integrate out the even spins;
the off-manifold fluctuations of the odd spins must be included
for consistency in the $1/N_c$ expansion.

The counterpart of the action (\ref{Srescaled}) for our problem
has an antiferromagnetic spin-spin interaction, with no $(-1)^\bn$
factors. We separate it into odd, even, and coupled terms,
\begin{eqnarray}
S &=& \frac{N_c}2\left(S^{\text{odd}} +
S^{\text{even}}+S^{AF}\right),
\nonumber   \\
S^{\text{odd}}&=&\int d\tau \sum_{\text{$\bn$ odd}}
\Tr\Lambda U_\bn^{\dag} \partial_{\tau} U_\bn, \nonumber    \\
S^{\text{even}}&=&\int d\tau \sum_{\text{$\bn$ even}}
\Tr\Lambda U_\bn^{\dag} \partial_{\tau} U_\bn, \nonumber    \\
S^{AF} &=& \int d\tau \sum_{\text{$\bn$ even}} \frac J2 \Tr
\sigma_\bn \sigma_\bn^o.
\end{eqnarray}
Here $\sigma_\bn^o = \sum_{\bmm(\bn)} \sigma_\bmm$, where
$\bmm(\bn)$ are the nearest neighbors of the even site $\bn$. We
expand the field on the even sites around $\sigma_\bn=\Lambda$ in
the manner of \Eq{sigmaL},
\begin{equation}
\sigma_\bn=\Lambda+L_\bn-\frac12L^2\Lambda\qquad\text{($\bn$
even),} \label{sigmaL2}
\end{equation}
while for the odd sites we write (see Appendix \ref{app:oddspins})
\begin{equation}
\sigma_\bn= U_\bn \left( \begin{array}{cc}
        \bm1_{2m-N}        & 0     \\
        0       & -\Lambda' - L' + \frac12 L'^2 \Lambda'
\end{array} \right) U_\bn^{\dag}\qquad\text{($\bn$ odd),}
\label{eq:Sig_1}
\end{equation}
with
\begin{equation}
\Lambda' = \left( \begin{array}{cc}
        \bm1_{N-m} &       0       \\
        0       &       -\bm1_{N-m}        \end{array} \right)
\end{equation}
and
\begin{equation}
U_\bn = \left( \begin{array}{cc}
        U^{(m)}_\bn     & 0     \\
        0       & \bm1_{N-m}       \end{array}     \right).
\end{equation}
$L_\bn$ describes the fluctuations of the even spins around their
classical value $\Lambda$. $U_\bn$ rotates the odd spins within
the manifold of their classical values, while $L'_\bn$ describes
their fluctuations outside that manifold. We further define
\begin{equation}
\sigma^{\text{cl}}_\bn=U_\bn \left( \begin{array}{cc}
        \bm1_{2m-N}        & 0     \\
        0       & -\Lambda'
\end{array} \right) U_\bn^{\dag}
=\left(\begin{array}{cc}
\sigma^{(m)}_\bn&0\\
0&\bm1_{N-m}
\end{array} \right),
\end{equation}
the classical field on the odd sites.

We leave $S^{\text{odd}}$ alone and expand $S^{\text{even}}$
and~$S^{AF}$ around the classical values of the fields,
\begin{eqnarray}
S^{\text{even}}&=& - \frac14 \int d\tau \sum_{\text{$\bn$ even}}
\Tr
\Lambda L_\bn \partial_\tau L_\bn,  \\
S^{AF} &=& S_0 + \frac{J}2 \int d\tau \sum_{\text{$\bn$ even}}
\left( \Tr L_\bn \bar\sigma_\bn -\frac12\Tr L_\bn^2 \Lambda
\bar\sigma_\bn
- \Tr L_\bn \bar L_\bn \right) \nonumber \\
&&\ +dJ \int d\tau \sum_{\text{$\bn$ odd}} \left(-\Tr\Lambda\tilde
L_\bn +\frac12\Tr\tilde L^2_\bn \right). \label{eq:SAF}
\end{eqnarray}
Here
\begin{equation}
\tilde L_\bn=U_\bn \left( \begin{array}{cc}
                0       &       0       \\
                0       &       L'_\bn\end{array} \right)
U^\dag_\bn
\label{tildeL}
\end{equation}
is the rotated fluctuation field on the odd sites, and the
Hermitian matrices $\bar\sigma_\bn$ and $\bar L_\bn$ are sums over
the odd neighbors of the even site $\bn$,
\begin{eqnarray}
\bar\sigma_\bn &=& \sum_{\bmm(\bn)} \sigma^{\text{cl}}_\bmm,    \\
\bar L_\bn     &=& \sum_{\bmm(\bn)} \tilde L_\bmm. \label{eq:Lo}
\end{eqnarray}
Since both $\bar\sigma_\bn$ and $\Lambda$ are block diagonal, the
first trace in each integral in \Eq{eq:SAF} is zero.

Now we organize the partition function as follows:
\begin{equation}
Z=\int \left(\prod_{\text{$\bn$ odd}}d\sigma_\bn\right)\, \exp
\left[-\frac{N_c}2 \left( S^{\text{odd}} + S_0 + \frac{dJ}2 \int
d\tau \sum_{\text{$\bn$ odd}}\Tr \tilde L^2_\bn \right) \right]
Z_{\text{even}}, \label{eq:Z}
\end{equation}
with
\begin{eqnarray}
Z_{\text{even}}&=&\int \left(\prod_{\text{$\bn$
even}}dL_\bn\right)\, \exp \left[-\frac{N_c}2 \int d\tau \right.
\nonumber\\
&&\left.\quad\times\sum_{\text{$\bn$ even}}\left( -\frac14 \Tr
\Lambda L_\bn \partial_\tau L_\bn -\frac J4 \Tr L_\bn^2 \Lambda
\bar\sigma_\bn -\frac J2 \Tr L_\bn \bar L_\bn \right)\right].
\end{eqnarray}
$Z_{\text{even}}$ is a product of decoupled single-site integrals.
Again we expand in the group algebra,
\begin{equation}
L_\bn = l_\bn^\eta M^\eta,
\end{equation}
where the sum is over the $2m(N-m)$ generators of $U(N)$ that are
not in $U(m)\times U(N-m)$. $\bar L_\bn$ can be expanded similarly
and we obtain the following form for the integral over the even
fields:
\begin{equation}
Z_{\text{even}} = \int Dl_\bn \exp \left[ -\frac{N_c}2 \int d\tau
\sum_\bn \left( l_\bn^\eta \mathcal M_\bn^{\eta\eta'}
l_\bn^{\eta'} +\frac J4 l^{\eta}_\bn \bar l^\eta_\bn \right)
\right],
\end{equation}
where
\begin{equation}
\mathcal M_\bn^{\eta \eta'}=-\frac18C^{\eta\eta'} \partial_{\tau}
+\frac J8 D^{\eta \eta'}_\bn.
\end{equation}
The matrix $C$ is the same as in \Eq{Cetaeta},
\begin{equation}
C^{\eta \eta'}=\Tr(\Lambda[M^\eta, M^{\eta'}]),
\end{equation}
while the new matrix $D$ varies with the site $\bn$ according to
the average $\bar\sigma_\bn$ of its neighboring spins,
\begin{equation}
D^{\eta\eta'}_\bn=-\Tr(\{M^{\eta},M^{\eta'}\}\Lambda\bar\sigma_\bn).
\label{Dmatrix}
\end{equation}
We study the two matrices in Appendix \ref{app:CD}. Diagonalizing
$C$ as before, we arrive at
\begin{equation}
Z_{\text{even}} = \prod_{\bn q}\left\{ \int Dl\,\exp -\frac{N_c}2
\int d\tau \left[ l^{q\dag} \hat\mathcal{M}_\bn l^q +\frac{J}4
\left( l^{q\dag}\bar{l}_\bn^q + \bar{l}_\bn^{q\dag}l^q \right)
\right] \right\},
\end{equation}
where $\hat\mathcal{M}_\bn$ is the $m\times m$ matrix
\begin{equation}
\hat\mathcal{M}_\bn = \frac14 \left( \partial_{\tau} + JE_\bn
\right). \label{eq:M}
\end{equation}
The quantities $l^q$ (and $\bar l^q$) for each $q=1,\ldots,N-m$
are complex $m$-component vectors; they are rotations of the
$2m(N-m)$ real components $l^\eta$ (and $\bar l^\eta$) into the
basis that diagonalizes $C$. The matrix $E_\bn$ is given by
\Eq{eq:E}; it carries the dependence on $\bar\sigma_\bn$.
Performing the gaussian integration we get
\begin{equation}
Z_{\text{even}} = \prod_{\bn q} \frac1{\Det \hat\mathcal{M}_\bn}
\exp \left[ \frac{N_c}2 \left( \frac J4 \right)^2 \int d\tau\,\bar
l^{q\dag}_\bn \hat\mathcal{M}_\bn^{-1} \bar l_\bn^q \right].
\label{eq:Zeven}
\end{equation}

Finally we separate the integral (\ref{eq:Z}) over the odd spins
into an integral over the classical field $\sigma_\bn^{\text{cl}}$
and an integral over the fluctuations around it.  We obtain
\begin{eqnarray}
Z&=&\int D\sigma_\bmm^{\text{cl}}\, \exp -\frac{N_c}2\left(S_0
+(N-m)\sum_{\bn}\Tr\log\hat\mathcal{M}_\bn \right) \nonumber \\
&& \times\int D\tilde l_\bmm \,\exp -\frac{N_c}2
\left\{S^{\text{odd}} + \int d\tau \left[\frac{dJ}2 \sum_{\bmm q}
|\tilde l^q_\bmm|^2 -\left(\frac J4 \right)^2\sum_{\bn q} \bar
l_\bn^{q\dag}\hat\mathcal{M}_\bn^{-1} \bar l_\bn^{q} \right]
\right\}. \label{eq:Z1}
\end{eqnarray}
Here $\bmm$ stands for an odd site, $\bn$ for an even one.

Equation~(\ref{eq:Z1}) gives an effective action for the classical
odd spins $\sigma_\bmm^{\text{cl}}$.
These enter the exponents through $\hat\mathcal{M}$ [via
Eqs.~(\ref{eq:E}) and~(\ref{eq:M})] and through
$\bar{l}^{q}_\bn$
[via Eqs.~(\ref{tildeL}) and~(\ref{eq:Lo})].
The action in the first exponent
is minimized when each matrix $E_\bn(\sigma_\bmm^{\text{cl}})$
has the largest number of zero
eigenvalues, each of which makes 
$\Tr\log\hat\mathcal{M}_\bn$ approach $-\infty$. It
is easy to check that $E_\bn$ has $2m-N$ zero eigenvalues (the maximal
number) when the $\sigma_\bmm^{\text{cl}}$ on {\em all} the odd
sites $\bmm(\bn)$ align, {\em i.e.,}
\begin{equation}
\sigma^{\rm cl}_\bmm = \sigma_0 \in U(m)/[U(2m-N)\times U(N-m)].
\label{eq:minimum}
\end{equation}
Moreover, when \Eq{eq:minimum} holds, all the $\tilde{l}_\bmm^q$'s
align parallel to each other and $\bar{l}_\bn^{q}$ is maximized;
also the eigenvalues of $\hat\mathcal{M}_\bn^{-1}$ are
maximized (to $+\infty$).
Thus the action in
the second exponent also has a minimum at
this point in configuration space.
These effects add up to an effective {\em ferromagnetic\/}
interaction among the $2d$ nearest neighbors $\bmm$ of each even
site $\bn$. This effective interaction will
align the classical spins on the odd sublattice to the same direction in
their submanifold, $U(m)/[U(2m-N)\times U(N-m)]$.

The divergences in the effective action have their origin in the fact
that the semiclassical corrections are calculated as gaussian integrals
in the even fluctuation fields $L_\bn$, and the coefficient matrix
$\mathcal M_\bn$ acquires zero eigenvalues.
The correct range of integration over $L_\bn$ is of course not infinite,
but rather the volume of the $U(N)/[U(m)\times U(N-m)]$ manifold.
This will regulate the divergences, but leave the effective action for
the odd spins attractive.

\section{Fluctuations on the odd sites \label{app:oddspins}}
\label{sec:App4}

In the classical analysis, the fields on the odd sites take values
in the sub-manifold $U(m)/[U(2m-N)\times U(N-m)]$ of the manifold
$U(N)/[U(m)\times U(N-m)]$. We denote these values
$\sigma^{\text{cl}}$,
\begin{equation}
\sigma^{\text{cl}}= \left( \begin{array}{cc}
    \sigma^{(m)} & 0    \\
    0   &   \bm1_{N-m}   \end{array}    \right). \label{sigmcl}
\end{equation}
Here
\begin{equation}
\sigma^{(m)} = U^{(m)} \Lambda^{(m)} U^{(m)\dag}. \label{eq:Sigm}
\end{equation}
with $U^{(m)} \in U(m)$, and
\begin{equation}
\Lambda^{(m)} = \left( \begin{array}{cc}
    \bm1_{2m-N} & \\
    0   & -\bm1_{N-m}   \end{array} \right).
\end{equation}

$\sigma^{(m)}$ contains $2(N-m)(2m-N)$ independent degrees of
freedom. Any $\sigma\in U(N)/[U(m)\times U(N-m)]$ can be written
as
\begin{equation}
\sigma=\left( \begin{array}{cc} \cos\left(2\sqrt{aa^{\dag}}\right)
& -a\,\frac{\displaystyle\sin\left(2\sqrt{a^{\dag}a}\right)}
    {\displaystyle\sqrt{a^{\dag}a}} \\[2pt]
-\frac{\displaystyle\sin\left(2\sqrt{a^{\dag}a}\right)}
    {\displaystyle\sqrt{a^{\dag}a}}\,a^{\dag}   &
-\cos\left(2\sqrt{a^{\dag}a}\right) \end{array} \right)
\label{eq:Sig_gen}
\end{equation}
[cf.~\Eq{eq:general_Q}], which coincides with \Eq{sigmcl} if
\begin{equation}
a =  U^{(m)}\left( \begin{array}{c}
    0   \\
    (\pi/2) \bm1_{N-m}  \end{array} \right).
\end{equation}
Recall that $a$ is an $m\times(N-m)$ matrix, so the zero block has
dimensions $(2m-N)\times(N-m)$.

We allow motion out of the sub-manifold by allowing $a$ to vary
further,
\begin{equation}
a =  U^{(m)}\left( \begin{array}{c}
        0       \\
        \bar{a} \end{array}     \right).
\end{equation}
The $2(N-m)^2$ degrees of freedom in $\bar a$ complement the
$2(N-m)(2m-N)$ degrees of freedom inherent in $U^{(m)}$ to give
$2m(N-m)$, the dimensionality of the entire $U(N)/[U(m)\times
U(N-m)]$ coset space.

Writing $\sigma$ with the generalized $a$ we have
\begin{equation}
\sigma= U \left( \begin{array}{ccc}
\bm1_{2m-N} &   0   &   0   \\[2pt]
0   & \cos\left(2\sqrt{\bar{a}\bar{a}^{\dag}}\right) &
-\bar{a}\,\frac{\displaystyle\sin\left(2\sqrt{\bar{a}^{\dag}\bar{a}}\right)}
    {\displaystyle\sqrt{\bar{a}^{\dag}\bar{a}}} \\[2pt]
0 &
-\frac{\displaystyle\sin\left(2\sqrt{\bar{a}^{\dag}\bar{a}}\right)}
{\displaystyle\sqrt{\bar{a}^{\dag}\bar{a}}}\,\bar a^\dag    &
-\cos\left(2\sqrt{\bar{a}^{\dag}\bar{a}}\right)
\end{array} \right)
    U^{\dag},
\end{equation}
with
\begin{equation}
U = \left( \begin{array}{cc}
        U^{(m)}     & 0     \\
        0       & \bm1_{N-m}       \end{array}     \right).
\end{equation}
We can also write this as
\begin{equation}
\sigma = U \left( \begin{array}{cc}
    \bm1_{2m-N} & 0 \\
    0   & \sigma^{[2(N-m)]}(\bar{a},\bar{a}^{\dag})
\end{array} \right) U^{\dag}.
\end{equation}
$\sigma^{[2(N-m)]}$ is a matrix in the manifold
$U(2(N-m))/[U(N-m)\times U(N-m)]$. Indeed for $\bar{a}=(\pi/2)
\bm1_{N-m}$, we have
\begin{equation}
\sigma^{[2(N-m)]} = \left( \begin{array}{cc}
    -\bm1_{N-m} &   0   \\
    0   &   \bm1_{N-m}  \end{array} \right) \equiv -\bar{\Lambda}.
\end{equation}
Since $U(2(N-m))/[U(N-m)\times U(N-m)]$ is a self-conjugate
manifold, its structure near $\sigma^{[2(N-m)]}=-\bar\Lambda$ is
the same as its structure near $\sigma^{[2(N-m)]}=\bar\Lambda$,
which corresponds to $\bar{a}=0$. Expanding $\sigma^{[2(N-m)]}$
around $-\bar\Lambda$ gives \Eq{eq:Sig_1}.

\end{document}